\documentclass[12pt]{article}
\usepackage{graphicx}
\usepackage{amsmath}
\usepackage{lipsum}
\usepackage{hyperref}
\usepackage{enumitem}
\usepackage{float}
\usepackage{enumitem}
\usepackage{geometry}
\title{\textbf{Distributedness based scheduling}}
\author{
 Paritosh Ranjan \\
  IBM  \\
  \texttt{paranjan@in.ibm.com} \\
  \and
 Surajit Majumder \\
  IBM  \\
  \texttt{surajit.majumder@ibm.com} \\
  \and
 Prodip Roy \\
  IBM  \\
  \texttt{prodipro@in.ibm.com} \\
  \and
 Bhuban Padhan \\
  IBM  \\
  \texttt{
bhubanpadhan@in.ibm.com}
}
\date{\today}

\begin{document}

\maketitle

\begin{abstract}
Efficient utilization of computing resources in a Kubernetes cluster is often constrained by the uneven distribution of pods with similar usage patterns. This paper presents a novel scheduling strategy designed to optimize the distributedness of Kubernetes resources based on their usage magnitude and patterns across CPU, memory, network, and storage. By categorizing resource usage into labels such as "cpu-high-spike" or "memory-medium-always," and applying these to deployed pods, the system calculates the variance—or distributedness factor—of similar resource types across cluster nodes. A lower variance indicates a more balanced distribution. The Kubernetes scheduler is enhanced to consider this factor during scheduling decisions, placing new pods on nodes that minimize resource clustering. Furthermore, the approach supports redistribution of existing pods through simulated scheduling to improve balance. This method is adaptable at the cluster, namespace, or application level and is integrated within the standard Kubernetes scheduler, providing a scalable, label-driven mechanism to improve overall resource efficiency in cloud-native environments.
\end{abstract}

\section{Introduction}

In Kubernetes-based environments, workloads—represented by pods—exhibit diverse patterns of resource consumption. These patterns vary across computing dimensions such as memory, CPU, network bandwidth, and storage usage, and may range from consistently high or low usage to sporadic spikes or gradual increases. Consequently, different pods place different demands on the underlying infrastructure in terms of both intensity and temporal characteristics. Current scheduling strategies often overlook the nuanced usage patterns and magnitudes of resource consumption, leading to suboptimal clustering of similar workloads on the same nodes. This can result in localized resource contention and inefficient utilization at the cluster level. To address this, it is essential to consider the distribution of pods with similar resource usage profiles across nodes. Specifically, pods exhibiting comparable consumption patterns and magnitudes for a given resource type should be evenly dispersed throughout the cluster to enhance overall resource efficiency and prevent hotspots in resource demand.

\section{Brief Description of the Invention}

This invention presents a novel scheduling approach for Kubernetes clusters aimed at maximizing resource efficiency by strategically distributing pods according to their unique resource usage patterns and intensities. The system assigns descriptive labels to pods, reflecting both the type (such as spike, gradual, or constant) and the level (low, medium, or high) of their consumption across CPU, memory, network, and storage resources. These labels inform the scheduler’s decisions, ensuring that pods with similar resource profiles are spread evenly across cluster nodes. By analyzing the variance—termed the distributedness factor—in the distribution of these labels among nodes, the scheduler can make smarter placement choices that help prevent resource bottlenecks. This methodology can also be applied retrospectively to rebalance existing workloads for improved cluster equilibrium. The approach is versatile, supporting deployment at various levels such as cluster-wide, by namespace, or by application, and it seamlessly integrates into Kubernetes’ scheduling framework without requiring major modifications to core components.

\section{Reduction to Practice}

The invention is realized by attaching standardized labels to Kubernetes pods that reflect their observed resource usage patterns and levels—including CPU, memory, network, and storage. These labels guide the Kubernetes scheduler, which employs variance-based analysis to identify the most balanced node for pod placement, ensuring that workloads with similar resource profiles are evenly distributed throughout the cluster. This approach can also be applied after initial deployment to rebalance existing pods, further enhancing overall resource utilization.

\section*{Steps}

\begin{enumerate}
    \item Labels are defined to represent different usage magnitudes and patterns for each type of computing resource.

    \begin{itemize}
        \item \texttt{memory-high-always}, \texttt{memory-medium-always}, \texttt{memory-low-always}
        \item \texttt{memory-high-spike}, \texttt{memory-medium-spike}, \texttt{memory-low-spike}
        \item \texttt{memory-high-gradual}, \texttt{memory-medium-gradual}, \texttt{memory-low-gradual}
        \item \texttt{cpu-high-always}, \texttt{cpu-medium-always}, \texttt{cpu-low-always}
        \item \texttt{cpu-high-spike}, \texttt{cpu-medium-spike}, \texttt{cpu-low-spike}
        \item \texttt{cpu-high-gradual}, \texttt{cpu-medium-gradual}, \texttt{cpu-low-gradual}
        \item \texttt{network-high-always}, \texttt{network-medium-always}, \texttt{network-low-always}
        \item \texttt{network-high-spike}, \texttt{network-medium-spike}, \texttt{network-low-spike}
        \item \texttt{network-high-gradual}, \texttt{network-medium-gradual}, \texttt{network-low-gradual}
        \item \texttt{storage-high-always}, \texttt{storage-medium-always}, \texttt{storage-low-always}
        \item \texttt{storage-high-spike}, \texttt{storage-medium-spike}, \texttt{storage-low-spike}
        \item \texttt{storage-high-gradual}, \texttt{storage-medium-gradual}, \texttt{storage-low-gradual}
    \end{itemize}

    \item These labels would be applied to the Kubernetes Resources deployed e.g., pods as per their computing resource’s usage.

    \item The cluster can distribute the application(i.e., calculate the distributedness) at three levels, the Resource definition should contain at what level any application has to be distributed:
    \begin{itemize}
        \item \textbf{Cluster level:} In this case, the system won't differentiate between different applications deployed in all namespaces of the cluster, and would distribute same or different applications only on the basis of their computing resource usage. Only the labels provided in the Resource definition would be used for this purpose e.g.,\texttt{cpu-low-spike}.
        
        \item \textbf{Namespace level:} In this case, the system won't differentiate between different applications deployed in a namespace, and the applications would be distributed at the namespace level. The label would be appended with namespace in this case e.g.,\texttt{cpu-low-spike-<namespace>}.
        
        \item \textbf{Application level:} In this case, the system would distribute the different replicas of an application deployed in a namespace. The label would be appended with namespace and application name in this case e.g.,\texttt{cpu-low-spike-<namespace>
        -<application-name>}.
    
    \end{itemize}

    \item These labels can be applied manually or by any automatic process to auto detect different labels based on their average or peak usage of any computing resource over a past time period.

    \item When a new scheduling request for a Kubernetes resource is received by the Kubernetes Master, then the magnitude and pattern usage label should already be available/applied to the Kubernetes resource.

    \item The Kubernetes scheduler would fetch the count of Kubernetes resources of that particular label deployed on each node of the Kubernetes cluster.
    
    \item This would provide a list of numbers where each number would represent the count of the number of Kubernetes resources of that particular label deployed on each node of the Kubernetes cluster.

    \item For example, suppose we have 10 nodes of the cluster with following IPs:

        \begin{enumerate}
            \item 10.220.45.89
            \item 10.220.45.56
            \item 10.220.45.2
            \item 10.220.45.148
            \item 10.220.45.34
            \item 10.46.7.204
            \item 10.46.7.168
            \item 10.46.7.8
            \item 10.46.7.10
            \item 10.46.7.129
        \end{enumerate}

    \item Following is the number of pods of label \textbf{“cpu-high”} deployed on each node for distrubtion-1 and distribution-2 i.e. two patterns of distribution:

    \textbf{Distribution 1}
    \begin{verbatim}
    10.220.45.89   - 2
    10.220.45.56   - 3
    10.220.45.2    - 4
    10.220.45.148  - 5
    10.220.45.34   - 1
    10.46.7.204    - 6
    10.46.7.168    - 1
    10.46.7.8      - 2
    10.46.7.10     - 8
    10.46.7.129    - 9
    \end{verbatim}

    \textbf{Distribution 2}
    \begin{verbatim}
    10.220.45.89   - 4
    10.220.45.56   - 5
    10.220.45.2    - 4
    10.220.45.148  - 4
    10.220.45.34   - 4
    10.46.7.204    - 4
    10.46.7.168    - 4
    10.46.7.8      - 4
    10.46.7.10     - 4
    10.46.7.129    - 4
    \end{verbatim}

    \item{Distributedness Factor:}  
    The distributedness factor of each numeric series is calculated. A higher distributedness factor indicates that data points in a set are spread out over a large range. 

    A low distributedness factor indicates that there is little spread or variation within the data set, and most values are very similar to the mean.

    \item{Calculation of distributedness factor:}  
    The distributedness factor is calculated by calculating the variance of the data series.For example, please see the variance of the two series for distribution-1 and distribution-2.Variance can be calculated using the following formula:
    It is calculated using variance:

    \[
    \text{Variance} = \frac{1}{n} \sum_{i=1}^{n}(x_i - \mu)^2
    \]

    \textbf{distribution-1}
    2, 3, 4, 5, 1, 6, 1, 2, 8, 9

    \textbf{variance} 
    $s^2 = 8.1$

    \textbf{distribution-2}
    4, 5, 4, 4, 4, 4, 4, 4, 4, 4

    \textbf{variance} 
    $s^2 = 0.1$
    
    We can see that the distribution-2 has lower distributedness factor than distribution-1, which means that distribution-2 is a better distribution than distribution-1.
    
    \item When any new pod or kubernetes resource is scheduled, then the scheduler will calculate the distributedness factor by theoretically putting the new instance to be deployed on every node and then pick the node for deployment where the distributedness factor is lowest after calculation.

    \item To redistribute the pods or other kubernetes resources of any existing cluster where the scheduling was done without considering the distributedness factor of the cluster, the system would simulate scheduling of pods or other kubernetes resources on each node one by one and calculate the distributedness factor for each simulated scheduling of pod or any other kubernetes resource. 
    
    \item The scheduling plan with the lowest distributedness factor would be selected.
    \item This system would be deployed in the “Scheduler” of the master node of existing Kubernetes architecture.
\end{enumerate}

\section{Advantages of the Invention}

The key advantages of the proposed invention are as follows:

\begin{itemize}
    \item \textbf{Enhanced Resource Allocation:} 
    The invention promotes even distribution of pods with similar resource consumption patterns and intensities across cluster nodes, helping to eliminate resource hotspots and boost overall efficiency.

    \item \textbf{Boosted Cluster Performance and Reliability:} 
    By preventing the grouping of resource-heavy pods on the same node, the system minimizes resource contention, resulting in steadier performance and greater system stability.

    \item \textbf{Intelligent, Variance-Driven Scheduling:} 
     Leveraging statistical variance (the distributedness factor), the scheduler gains a data-driven, effective method for making placement decisions, significantly increasing its intelligence and effectiveness.

    \item \textbf{Effortless Kubernetes Integration:} 
    The solution fits seamlessly into the existing Kubernetes scheduling framework, requiring no substantial modifications to cluster infrastructure and maintaining full compatibility with standard operational workflows.

    \item \textbf{Flexible and Scalable Granularity:} 
    The system supports customizable distribution policies at the cluster, namespace, and application levels, making it highly adaptable to diverse organizational structures and deployment scenarios.
\end{itemize}

\section{Conclusion}
This invention presents an innovative and structured method for optimizing Kubernetes resource scheduling by intelligently distributing pods according to their resource consumption levels and behavioral patterns. Through the use of descriptive labels that encapsulate resource characteristics, and by leveraging statistical variance (the distributedness factor) to inform placement decisions, the system greatly improves resource utilization, cluster performance, and operational reliability. It integrates smoothly with the existing Kubernetes architecture and supports flexible workload distribution at the cluster, namespace, and application levels. Additionally, the solution facilitates both real-time scheduling enhancements and retrospective workload rebalancing, providing a scalable and practical answer to common inefficiencies in container orchestration. As a result, this approach marks a significant step forward in intelligent workload placement and infrastructure optimization for cloud-native environments.

\section{Acknowledgment}

We would like to express our sincere gratitude to all individuals and organizations who have contributed to the success of this research. We acknowledge the invaluable support from the IBM team, whose resources and expertise have greatly enhanced this project.
Special thanks to Prodip Roy (Program Manager IBM) for their insightful feedback, guidance, and encouragement throughout the development of this work.
\section{References}
\renewcommand\refname{}

\end{document}